\documentclass[10pt, conference]{IEEEtran}
\IEEEoverridecommandlockouts
\usepackage{cite}
\usepackage[cmex10]{amsmath}
\usepackage{amssymb,amsfonts}
\usepackage{algorithmic}
\usepackage[pdftex]{graphicx}
\usepackage{textcomp}
\usepackage{xcolor}
\usepackage[colorlinks=true,urlcolor=black,linkcolor=black,citecolor=black]{hyperref}
\usepackage{booktabs}
\usepackage{tabularx}
\usepackage{balance}
\usepackage{multirow}
\usepackage{multicol}
\usepackage{xspace}
\usepackage{microtype}
\usepackage[capitalize]{cleveref}
\crefname{section}{Sect.}{sections}
\Crefname{section}{Section}{Sections}

\definecolor{darkolivegreen}{rgb}{0.33, 0.42, 0.18}

\usepackage{ifthen}
\newboolean{showcomments}
\setboolean{showcomments}{false} 
\ifthenelse{\boolean{showcomments}}
  {\newcommand{\nb}[2]{
    \fcolorbox{gray}{yellow}{\bfseries\sffamily\scriptsize#1}
    {\sf\small$\blacktriangleright$\textit{#2}$\blacktriangleleft$}
  }
  
  }
  {\newcommand{\nb}[2]{}
  
  }

\newcommand{\ric}[1]{\nb{Ricardo}{\footnotesize #1}}

\newcommand{\raz}[1]{\nb{Razan}{\footnotesize #1}}
\newcommand{\tb}[1]{\nb{Thorsten}{\footnotesize #1}}

\newcommand{\one}[1]{\nb{Reviewer 1}{\footnotesize #1}}
\newcommand{\two}[1]{\nb{Reviewer 2}{\footnotesize #1}}
\newcommand{\thr}[1]{\nb{Reviewer 3}{\footnotesize #1}}
\newcommand{\four}[1]{\nb{Reviewer 4}{\footnotesize #1}}

\begin{document}

\title{\textit{Towards Mapping Control Theory and Software Engineering Properties using Specification Patterns}}



\author{
\IEEEauthorblockN{\hspace{40pt}Ricardo Caldas and Razan Ghzouli}
\IEEEauthorblockA{
    \textit{\hspace{40pt}Chalmers $|$ U. of Gothenburg, Sweden}
  }
 \and
 \and
 \IEEEauthorblockN{Alessandro V. Papadopoulos}
 \IEEEauthorblockA{
 \textit{M{\"a}lardalen University, Sweden}
 }
 \and
 \IEEEauthorblockN{\hspace{50pt}}
 \IEEEauthorblockA{\hspace{50pt}
 \textit{}
 }
 \and
 \IEEEauthorblockN{\hspace{30pt} Patrizio Pelliccione}
 \IEEEauthorblockA{
 \textit{\hspace{30pt}Chalmers $|$ U. of Gothenburg, Sweden}\\ 
 \textit{\hspace{30pt}Gran Sasso Science Institute, Italy}
 }
 \and
 \IEEEauthorblockN{Danny Weyns}
 \IEEEauthorblockA{
 \textit{KU Leuven, Belgium} \\
 \textit{Linnaeus U., Sweden}\\
 }
 \and
 \IEEEauthorblockN{Thorsten Berger}
 \IEEEauthorblockA{
 \textit{Chalmers $|$ U. of Gothenburg, Sweden}\\
 \textit{Ruhr University Bochum, Germany}
 }
}

\newtheorem{definition}{Definition}


\def\BibTeX{{\rm B\kern-.05em{\sc i\kern-.025em b}\kern-.08em
    T\kern-.1667em\lower.7ex\hbox{E}\kern-.125emX}}

\makeatletter
\newcommand{\linebreakand}{%
  \end{@IEEEauthorhalign}
  \hfill\mbox{}\par
  \mbox{}\hfill\begin{@IEEEauthorhalign}
}

\makeatletter
\def\ps@IEEEtitlepagestyle{
        \def\@oddfoot{\mycopyrightnotice}
        \def\@evenfoot{}
}
\def\mycopyrightnotice{
        {\footnotesize
                \begin{minipage}{\textwidth}
                        \centering
                        \textcopyright~2021 IEEE.  Personal use of this material is
permitted.  Permission from IEEE must be obtained for all other uses, in
any current or future media, including reprinting/republishing this
material for advertising or promotional purposes, creating new
collective works, for resale or redistribution to servers or lists, or
reuse of any copyrighted component of this work in other works.
                \end{minipage}
        }
}

\maketitle



\thr{The paper is submitted under the vision category, which the organizers have asked to judge primarily on non-technical aspects, rather on the potential impact the proposed ideas would have on the field. Unfortunately, this paper doesn’t do enough of this. The only proposed idea is that rather straightforward definition of stability using LTL. There is no discussion on alternatives options considered, how this would compare with the mechanics of control theory (such Eigenvalues in the negative half-plane or within a unit-circle), what such qualities would actually mean for the field, or any attempt to define further such qualities in LTL (or any other SE-oriented representation).}
\four{I also think there's a larger point to be made about the differences between traditional and cyber-physical systems, in that the latter often have "moving targets" that change the goals of correctness even during the recovery from previous disruptions.}
\begin{abstract}
    
A traditional approach to realize self-adaptation in software engineering (SE) is by means of feedback loops. The goals of the system can be specified as formal properties that are verified against models of the system. On the other hand, control theory (CT) provides a well-established foundation for designing feedback loop systems and providing guarantees for essential properties, such as stability, settling time, and steady state error. Currently, it is an open question whether and how traditional SE approaches to self-adaptation consider properties from CT. Answering this question is challenging given the principle differences in representing properties in both fields. 
In this paper, we take a first step to answer this question. We follow a bottom up approach where we specify a control design (in Simulink) for a case inspired by Scuderia Ferrari (F1) and provide evidence for stability and safety. The design is then transferred into code (in C) that is further optimized. Next, we define properties that enable verifying whether the control properties still hold at code level. Then, we consolidate the solution by mapping the properties in both worlds using specification patterns as common language and we verify the correctness of this mapping. The mapping offers a reusable artifact to solve similar problems. Finally, we outline opportunities for future work, particularly to refine and extend the mapping and investigate how it can improve the engineering of self-adaptive systems for both SE and CT engineers.
    
\end{abstract}
    
\section{Introduction}
\label{sec:intro}
\looseness=-1
\noindent Providing evidence that a self-adaptive system behaves according to the stakeholders' requirements is challenging\cite{lemos:2017:assurance:challenges,weyns:2019:perpetual,weyns2020book}, especially when the system operates in uncertain environments. A potential remedy is to exploit principles from control theory (CT) to engineer self-adaptive systems~\cite{hellerstein:2004:feedback}. Control theory provides a mathematical framework for designing and analyzing dynamic systems, offering a formal basis to provide guarantees for control properties such as stability, overshoot, and settling time. For computing systems, CT has primarily been used to manage low-level resources, such as CPU cycles, communication bandwidth, and hardware~\cite{1657688}. Recently, there has been an increasing interest in applying the mathematical framework of CT to
 control software elements~\cite{FilieriHM14,shevtsov:2018:litrev,maggio:2017,10.1145-3105748,maggio:2020:control} as well. 

\looseness=-1
A common approach in software engineering (SE) to provide guarantees for a self-adaptive system is to test it against its requirements, or formally, by specifying formal properties\footnote{A property expresses what a system should or should not do, or how a system should or should not behave. To analyze a property, we need a property  specification~\cite{lamport2015builds} that for instance can be verified against a model. In this paper, the term \textit{property} refers to formally specified properties.} that are verified against models of the self-adaptive system~\cite{Brun2009,ActivFORMS,calinescu:2017:entrust}. 
Such formal approaches work case by case using tools such as model checkers. On the other hand, CT-based solutions provide \textit{guarantees by design}---that is, controller models properly designed according to the mathematical principles of CT satisfy the target CT properties. Yet, understanding how, and to what extent, traditional SE  approaches to self-adaptation consider and comply with properties from CT is challenging given the principle differences between the two paradigms. 

\looseness=-1
In this paper, we take a first step towards determining whether and how traditional SE approaches to self-adaptation consider properties from  CT. 
The relationship between SE approaches to self-adaptation and CT has been investigated from different angles. One line of research---reflected in Brun et al.~\cite{Brun2009} and Filieri et al.~\cite{filieri:2015:ct-meets-se}, among others---determines the mapping of the elements of a CT feedback loop design to the elements of the MAPE-K architecture~\cite{kephart2003vision}. Another line of research---reflected in Shevtsov et al.~\cite{shevtsov:2017:handling} and Caldas et al.~\cite{caldas:2020:hybrid}, among others---studies the synthesis of controllers for correct and efficient adaptation based on control theory. Recently, C\'amara et al.~\cite{camara:2020:bridging} made a step forward in bridging the gap by proposing a mapping between CT properties and self-adaptive systems properties relying on a common language. We stand on their shoulders to investigate the foundational issue of whether and to what extent properties of traditional SE approaches to self-adaptation consider properties from CT.

\looseness=-1
Inspired by a concrete case---the Simulink-model-based spe\-ci\-fication of a control design by Scuderia Ferrari (F1)~\cite{ferrari:2015}---the engineering process used in our work
includes adaptation goal identification, CT design, controller implementation and integration, and finally validating the implementation against the goals. Once the goals include CT properties, the challenge arises to check whether the implementation satisfies those CT properties. Here, the interplay between SE and CT takes place. For instance, suppose that the software engineers introduce a fault in the software when optimizing C code, and suppose that this fault induces behavior that violates the requirement of keeping a safe distance from the vehicle ahead, previously guaranteed by the controller design.
How can the test engineers verify that the safety requirements considered by the control engineers in the design still hold? To enable engineers to check CT properties on the implementation, we envision a mapping between CT properties and SE properties (here formulated in LTL). A formulation of CT properties in some kind of temporal logic might enable, for instance, the use of model checkers (e.g., DIVINE~\cite{BBK+17} or Uppaal~\cite{dimacs95}), monitoring techniques (e.g., Larva~\cite{Larva} or PREDIMO~\cite{zhang2018automatic}), and model-based testing~\cite{MBTsurvey}. To that end, we perform initial steps towards determining the mapping by exploiting  so-called property specification patterns~\cite{dwyer:1999:patterns,autili:2015:aligning}. We show the overall engineering process and mapping through an example of adaptive cruise control. 
We also check the defined mapping between CT and SE properties by investigating whether the LTL formula (SE property) presents the same behaviors that can be observed at the CT simulation level (CT property). To that end, we use the model checker DIVINE~\cite{BBK+17}. The results of the validation are promising, indicating feasibility and effectiveness of identified mappings between CT and SE properties. We provide an online replication package~\cite{appendix:online}, including our artifacts (e.g., Simulink model) and further details.

\section{Background and Motivation} 
\label{sec:problem}

\noindent
\textbf{Engineering Process.} Engineering a controller to realise self-adaptation typically involves an iterative process where control experts work separately from software engineers~\cite{liu:2004}. Such a process usually comprises six steps: (1) identify the adaptation goals, (2) identify the knobs (a.k.a., configuration options or calibration parameters), (3) devise the system model, (4) design the controller, (5) implement and integrate the controller, and (6) test and validate the system~\cite{filieri:2015:ct-meets-se}.
Especially step (5), the implementation and integration of controllers into a larger system, is considered a challenging task~\cite{hellerstein:2009:keynote}. In principle, engineers follow a model-driven approach~\cite{neis:2019:power-plants, schaefer:2021:future}, relying on a variety of tools and modeling languages (e.g., (MATLAB/Simulink, Modelica, SCADE), where the control model is transformed into hardware- and environment-specific models and eventually into source code. Still the generated code, as we learned from the Ferrari case, needs to be further optimized and customized before it can be integrated into a larger system. This requires extensive manual modification of the code~\cite{krikava:2018:control-integration}---an error-prone activity that may introduce bugs that are hard to spot without proper verification techniques. In addition, when the controller becomes part of a larger system, it is also influenced by it, further questioning whether the necessary CT properties still hold. Apparently, verifying code requires SE verification techniques, but to what extent the system does (or can) address CT properties is an open question.

\vspace{3pt}\noindent
\textbf{CT versus SE Properties.}
\looseness=-1
Mapping and comparing SE and CT properties is challenging. The former are typically formulated over a model of the system (often graph-based, such as a finite or a B\"uchi automaton) and expressed in a formal language, such as in temporal logics or in a process algebra. The latter are completely defined in algebraic formulations. While challenging, we believe that a mapping between the properties of the two worlds might be facilitated by expressing them in a common notation. Current approaches, such as C\'amara et al.~\cite{camara:2020:bridging}, use a formal language as common notation. Although flexible, specifying properties using a common formal language can be time-consuming and prone to errors.

\vspace{3pt}\noindent
\textbf{Process and Properties at Ferrari.}
We further illustrate this problem based on the process followed by Scuderia Ferrari~\cite{ferrari:2015}.
The manufacturer develops software for high-performance racing cars under tight one-week sprints. A typical requirement to be addressed from the control software involves four engineers: control experts, software developers, testers, and track engineers. At the highest level of development, control experts specify controller behavior using block diagrams aiming to guarantee the required control properties (e.g., stability, settling time, overshoot). The controller specifications are transformed into lower-level C code, where software engineers apply customizations and optimizations. The modified code is then thoroughly tested using software-in-the-loop (SIL) or hardware-in-the-loop (HIL) techniques. At the testing level, testers and control experts monitor the simulated behavior to guarantee that it behaves according to the design, i.e., that the car behavior complies with the control properties. Finally, the control software is deployed. At runtime, the track application engineers constantly monitor, elicit bug-fixing or improvement requirements, and in emergency cases perform hot-fixes to the running vehicle. The pipeline is executed in 7 days of intense work, from which half is concerned with specification, two with coding, one with testing, and the last three with validation.

\vspace{3pt}\noindent
\textbf{Mapping Properties using Specification Patterns.}
As motivated above, understanding the extent to which properties considered by traditional SE approaches to self-adaptation cover fundamental properties of CT urges for a mapping between the two. To this end, we propose a technique that uses patterns to map CT properties to SE properties.
Inspired by the process and details of how Ferrari engineers controllers, in this short paper, we demonstrate the feasibility and effectiveness of mapping CT and SE properties. Specifically, we design an example control model in Simulink with guarantees for CT properties. The model is then transformed into C code and is subject to modifications. Next, we define properties and verify whether the control properties still hold at code level. To consolidate the solution, we map the properties in both worlds using specification patterns as a common language, and we verify the correctness of this mapping.
In the next \cref{sec:design,sec:se_prop,sec:check}
we explain this process using a running example.

\section{System Model Design}\label{sec:design}

\begin{figure*}[htb!]
 	\centering
 	\includegraphics[
 	width=0.9\textwidth]{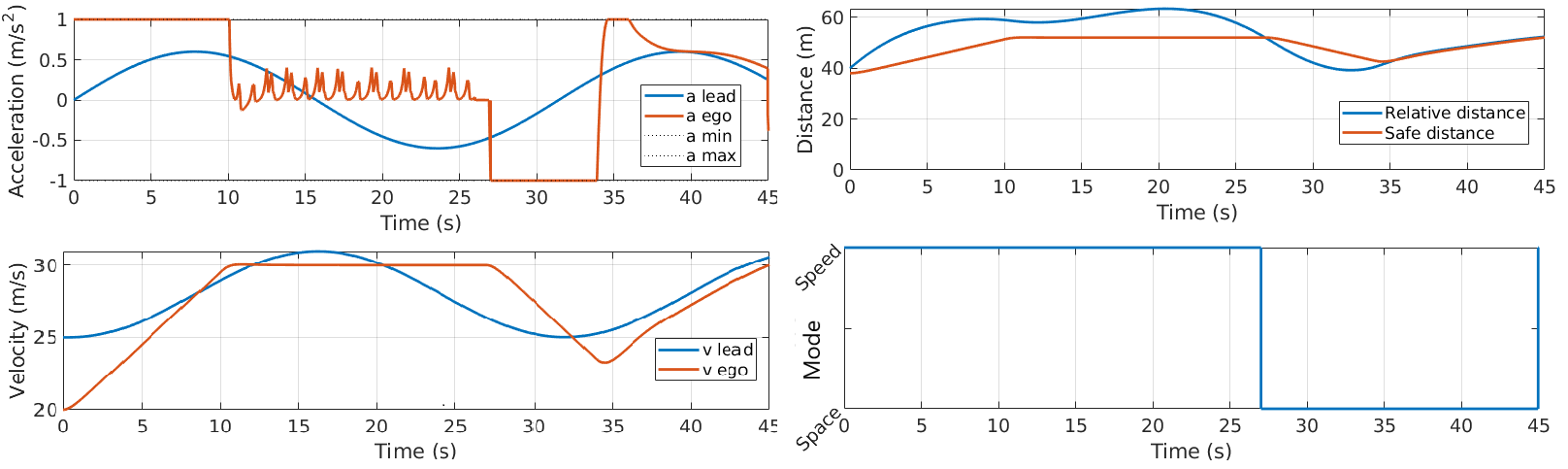}
     \caption{Simulation excerpt of the designed CT model. Left: relation between acceleration, and leading and ego vehicles' speed. Right: relation between  $D_{rel}$ and $D_{safe}$, and the switching behavior between space mode and speed mode controllers.}
 	\label{fig:perodicSimulation}
 	\vspace{-6mm}
 	
\end{figure*}


\noindent
The process starts with the design of a control model that complies to specific requirements. Using Simulink, we designed an adaptive cruise control (ACC), a driver assistance system, for a vehicle (ego vehicle). We modeled a scenario where the ego vehicle is provided with an ACC that automatically tracks a set velocity and adjusts the ego vehicle speed to maintain a safe distance from a preceding vehicle (lead vehicle)~\cite{marsden:2001:towards}. The ACC system has two operating modes: (1) Speed control where the ego vehicle follows a given speed ($V_\textit{set}$) and (2) spacing control to keep a safe distance from the lead vehicle ($D_\textit{safe}$). 

We used a model-based approach to design two proportional–integral–derivative (PID) controllers, speed mode PID and space mode PID, that guarantee the requirement: 
``If the relative distance between the two vehicles ($D_\textit{rel}$) is less than a safe distance ($D_\textit{safe}$), the controller of the ego vehicle should adjust its speed ($V_\textit{ego}$) for $D_\textit{rel}$ to become greater than $D_\textit{safe}$, otherwise follow a given  velocity ($V_\textit{set}$).'' In other words, if the safety of the vehicle is breached, the controller should adjust the vehicle's speed to maintain a safe distance. The relative distance ($D_\textit{rel}$) is the difference between the ego vehicle's position ($x_\textit{ego}$) and the lead vehicle's position ($x_\textit{lead}$), see \cref{eq:drel}. The safe distance is a function of the ego vehicle velocity, \cref{eq:dsafe}, where $D_{\textit{default }}$ is the standstill default spacing and $T_\textit{gap} \times V_\textit{ego}$ is the gap between the vehicles. $T_\textit{gap}$ is chosen to enable the ego vehicle to break without crashing the leading vehicle \cite{lin2009effects}. 

\begin{equation}
    \label{eq:drel}
    D_\textit{rel} = x_\textit{lead} -x_\textit{ego}
\end{equation}
\begin{equation}
    \label{eq:dsafe}
    D_\textit{safe}=D_\textit{default} + T_\textit{gap} \times V_\textit{ego}
\end{equation}

The switching between the two PIDs depends on a velocity error ($e_v$) and a distance error ($e_d$), see \cref{eq:switch}. We have used a state-based notation that produces the activation signal (1 for speed mode and -1 for space mode) followed by a switch to implement the mode switching in Simulink. With the switch the two PIDs do not operate simultaneously, satisfying the safety requirements.

\noindent\one{Equation (3) is unclear to me. In particular, the two conditions do not seem to be opposite to one another, so that Mode could be simultaneously both 1 and -1 (or neither).}
\two{First, their ACC system has two operating modes: (1) Speed control where the ego vehicle follow a set speed (Vset) and (2) spacing control to keep a safe distance from the lead vehicle (Dsafe).  In Figure 1, “The right figures shows the relation between Drel and Dsafe, and the switching behavior between space mode and speed mode controllers” and shows that switching behavior as instantaneous. As defined, it is clear why one would not have these two PIDs operating simultaneously, but are there other cases where one would have PIDs operating at the same time and how does that change perhaps their approach from the CT side?  Also is it a reasonable assumption in this case to have the PIDs switch behaviors instantaneously or is that just a nice first assumption to work out this approach?} \raz{addressed except for the last part of reviwer 2. I think it is maybe a case to check in the extension. For now commenting on this part will lead to a discussion about our approach applicability in different scenarios, which maybe an overkill for this version. what do you think? }
\begin{equation}
\label{eq:switch}
    \textit{Mode} = \begin{cases}
    1 \text{ speed mode} & \text{if } e_v = V_\textit{set} - V_\textit{ego} < 0\\
    -1  \text{ space mode} & \text{if } e_d = D_\textit{safe} - D_\textit{rel} >0
    \end{cases}
\end{equation}

\Cref{fig:perodicSimulation} shows an excerpt of the simulation showing the behavior of the designed controllers for 45 sec. In this setting, we simulated a scenario where the speed of the lead vehicle varies in time according to a sine wave resulting in changes in the distance between the two vehicles. The ACC switches between the space controller and the speed controller to keep the ego vehicle safe. At the moment 27 sec a breach in the safety requirement occurs resulting in the ACC reacting to keep safety. The reaction must lead into a state where the system settles around an equilibrium point. If this happens, the system is said to be stable.\footnote{There are techniques for the analysis of the stability of hybrid systems based on the physical model of the system, see~\cite{goedel:2012:stability}, but the description of such techniques is beyond the scope of the paper.} 

\section{Modeling Stability as a SE Property}\label{sec:se_prop}

\noindent
Guaranteeing that the system is stable at code-level (as illustrated in \cref{fig:perodicSimulation}), requires that the property is modeled in a formal language such as a graph-based formulation, temporal logic, or process algebra. In this section we explain how we modeled stability using linear temporal logic.

A control system is {\it stable} even if the error $e(t)$ is not converging to zero, but the error is bounded. More specifically, in control terms, if the initial value of the system output is ``close'' to the equilibrium value, then the evolution over time of the output of the system will be bounded~\cite{camara:2020:bridging}.

\begin{figure}[htb!]
  \centering
 	\includegraphics[width=0.90\columnwidth]{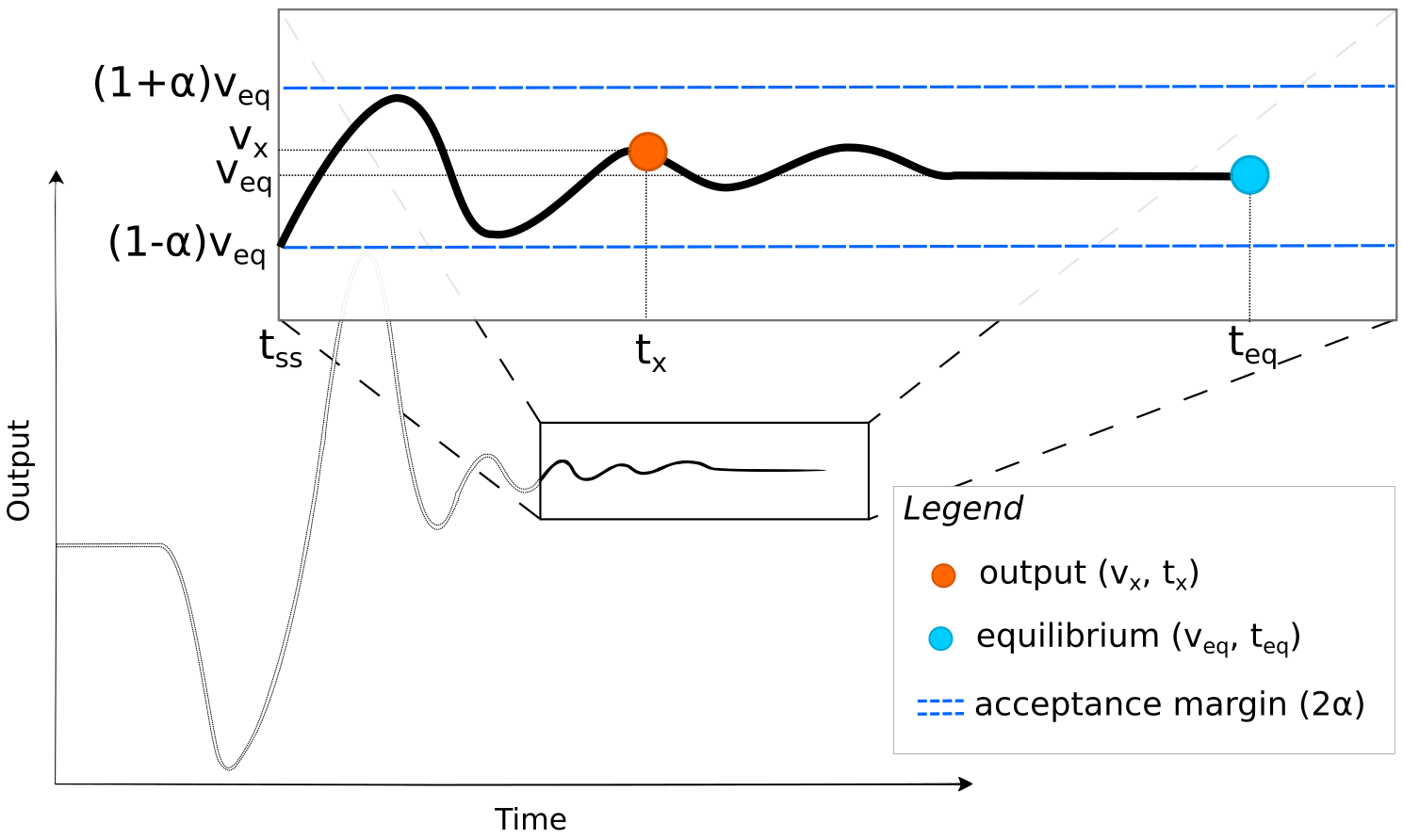}
 	\caption{Illustration of the CT property Stability for a step input.}
 	\label{fig:stabmod}
\end{figure}

Importantly, at some point the system must reach a so-called steady-state, where the signal is bounded. Therefore, we formalize steady-state beforehand. Let us suppose that a system is stimulated by a unitary step. Given the current state of the system's response output ($v_x$), the equilibrium value ($v_{eq}$) estimated with the output samples and the acceptance margin ($\alpha$) for convergence. With this in mind we formulate the steady-state condition (SS).

\begin{equation}
  \label{eq:ss}
  SS \equiv |v_{x} - v_{eq}| \leq \alpha
\end{equation}

\Cref{eq:ss} asserts that the steady-state condition holds when the distance between the current value and the equilibrium value is bounded by the acceptance margin.

\two{although the stability arguments are quite clear, it wasn’t as clear what was being defined as transients or how they are handled.  One definition used in the paper is “Transient is the period that ends when the response reaches the steady-state.”  That might work in this first case, but what of later problems where the steady state is never reached.  Would a better definition of transient maybe be define the time it takes to get near enough to the steady state and stay near enough  as the transient?}

The stability property is defined by the ability to reach and stay at steady-state. The output measurement of stable systems converges to steady-state if the system is `excited' by an input. \Cref{fig:stabmod} illustrates the response to a step input where the set of output values ($v_{x}$) is within a band defined by the acceptance margin ($\alpha$). The acceptance margin is a limiting value relative to the equilibrium point ($v_{eq}$).

How to formulate the CT stability property in LTL? We exploit specification patterns as a common language to create the mapping between CT and SE properties. A specification pattern~\cite{dwyer:1999:patterns, autili:2015:aligning} is defined as a tuple $<$\textit{Scope, Pattern}$>$. A Scope determines the \textit{extent of a program execution over which the pattern holds}\footnote{\url{https://matthewbdwyer.github.io/psp/patterns/scopes.html}}. Examples of scopes are Globally (entire execution trace) and After X (execution trace after a state/event X). A Pattern describes a \textit{generalized recurring system attribute}~\cite{autili:2015:aligning}. Examples of patterns are Universality (a property that always holds) and Existence (a property that eventually holds). A property can be specified by one specification pattern or a composition of multiple patterns using a nesting operator. For instance, the Universality and the Existence patterns might be composed under the Globally scope to obtain the Globally, Universality Existence (i.e., $\square \Diamond P$).

We rely on a syntactical comparison of CT and SE properties to map them using specification patterns. In our running example, we analyze the stability property from the two viewpoints. Stability as a CT property is represented by a feedback loop in which the ego's actual position is fed back into the controller. The ego's position is used to calculate the relative difference between vehicles and determine which acceleration will restore ego's safe position. In other words, stability determines the ability of the ego vehicle to restore and maintain itself in a wanted state for undetermined time.

\begin{table}[hb!]
 \centering
 \begin{tabular}{ll}
 \cline{1-2}
 \multicolumn{1}{|l|}{\tiny{Globally Untimed Existence Universality}}& \multicolumn{1}{l|}{} \\ 
 \cline{1-1}
 \multicolumn{2}{|l|}{\begin{tabular}[c]{@{}l@{}}\\\textbf{Globally eventually always} $\text{SS}$ \textbf{holds.} \\ \end{tabular}} \\
 \cline{1-2}
 \end{tabular}
\end{table}

We formalize the Stability property by employing the specification pattern Globally Existence, which aims at describing that events/states eventually holds, nested with Globally Universality specifying the case that the events/states always hold. 
Syntactically, the Globally Existence Universality coincides with the stability property. The Globally Existence Universality is represented in temporal logic as shown in \cref{eq:t}.

\one{There is something that doesn't feel quite right to me in (5). The formula $\Diamond \square SS$ is true if eventually, $SS$ becomes true and stays so forever. Doesn't this already capture stability? What is the rest of the formula adding? It seems to me that actually (5) is equivalent to $\Diamond \square SS$: in particular, if $SS$ does not become true and stays so forever, there are infinitely many points for which $\neg SS$ holds, and for none of them the parenthesised formula is true; making the whole formula false}

\begin{equation}
  \label{eq:t}
  \textit{Stability} \equiv \Diamond (\square SS) 
\end{equation}

In our running example, the ego vehicle must restore its position beyond the safety distance whenever the ego vehicle gets too close to the leading vehicle and maintains the safe distance. Therefore, we defined the steady-state as a function of \cref{eq:drel,eq:dsafe}, where $v_x = D_{\text{rel}}$, $v_{\text{eq}} = D_{\text{safe}}$, and $\alpha = 0.05 \times D_{\text{safe}}$.
Lastly, stability in the running example's context can be formulated as shown in \cref{eq:stab_safe}.

\begin{equation}
  \label{eq:stab_safe}
  \textit{Stability} \equiv \Diamond (\square (D_{\text{rel}} - D_{\text{safe}} > \alpha) ) 
\end{equation}

It is important to note the absence of the absolute operation around $D_{\text{rel}} - D_{\text{safe}}$. In this specific case, when the distance between ego and the leading car is greater than $\alpha = 0.05 \times D_{\text{safe}}$ the safety requirement is not violated and the system is considered stable.
\section{Checking the Mapping} \label{sec:check} 
\noindent
In this section, we use model checking to increase the confidence about the correctness of the mapping\footnote{Instructions on how to replicate our experiments as well as the technicalities of the model checking process are available in our online appendix\cite{appendix:online}.}.
\tb{clarify formulation: a property cannnot have behavior, a property is defined and either holds or not, what does that mean?}
To that end, we model check the SE property with three different scenarios in which we can attest whether the requirement is satisfied or not by visualizing the CT simulation output. Thus, we check the correctness of the mapping by comparing whether the LTL formula (SE property) holds in comparison with the expected result observed in the CT simulation level (CT property). As exemplified by Figure~\ref{fig:mcheck}, in simulation, we capture the former three scenarios behavior using Simulink. We auto-generate the C++ code for the System model using Simulink Coder\footnote{\url{https://se.mathworks.com/products/simulink-coder.html}} for each scenario. Using DIVINE 4\footnote{\url{https://divine.fi.muni.cz/index.html}} we assure the correctness of our mapping by feeding both the generated B\"uchi Automata in never claim representation\footnote{The B\"uchi Automata in never claim was manually encoded within DIVINE.}
(SE property) and the generated C++ code (CT property behavior) to the model checker. 

\begin{figure}[tb!]
  \centering
 	\includegraphics[width=0.75\columnwidth]{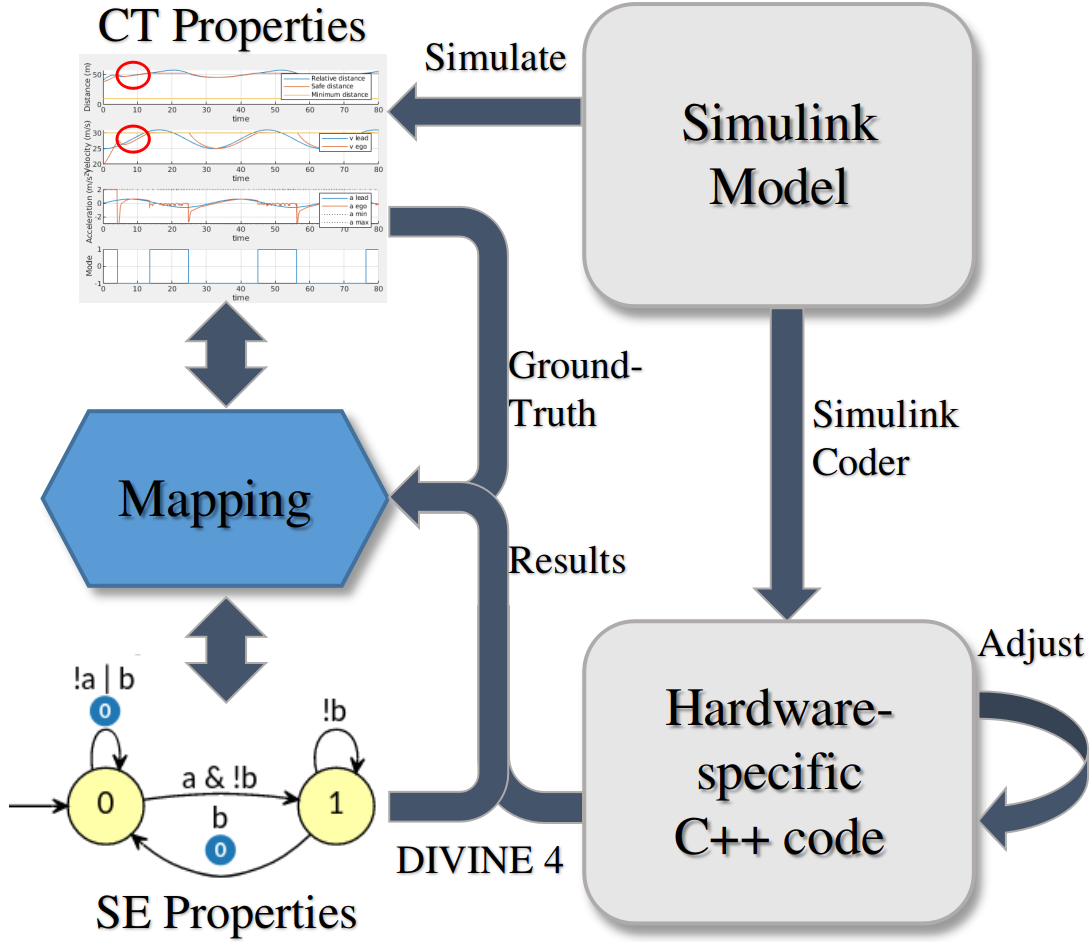}
 	\caption{Using DIVINE 4 to get confidence on the CT-SE properties mapping.}
 	\label{fig:mcheck}
\end{figure}

\begin{table}[h] 
\caption{Scenarios}
\begin{tabularx}{\columnwidth}{cccccX}

        \textbf{ID} & \textbf{$v_0$ ego} & \textbf{$v_0$ lead} & \textbf{$x_0$ ego} & \textbf{$x_0$ lead} & \textbf{Description} \\
        \midrule
        Case 1  & 10 km/h & 30 km/h & 10 m & 50 m & Ego is always at a safe distance. \\ 
        \midrule
        Case 2  & 20 km/h & 25 km/h & 3 m & 5 m  & Ego recovers from unsafe distance.\\   
        \midrule
        Case 3  & 40 km/h & 15 km/h & 10 m & 20 m & Ego cannot recover from unsafe distance. \\
        \bottomrule\vspace{-15pt}
		\label{tab:scenario}
\end{tabularx}
\end{table}

The scenarios in \cref{tab:scenario} are tailored to show whether there is a semantic equivalence between safety at CT level and SE level. Therefore, they need to explore the different behavior that might occur during  system execution. To generate the scenarios, we changed the initial settings of the experiment with different vehicle's starting positions ($x_0$) and starting speeds ($v_0$). 

We ran the simulation and model checked all scenarios on a Ubuntu 18.04, processor Intel(R) Core(TM) i7-8665U CPU @ 1.90GHz, and 32GB memory. All the scenarios returned the expected result, the Stability property (\cref{eq:stab_safe}) holds for cases 1 and 2 but not for case 3, see \cref{tab:results}.

\begin{table}[htb!] 
\caption{Model checking results}
\centering
\begin{tabularx}{0.9\columnwidth}{ccccc}
        \textbf{ID}& Mem. Used & Exec. Time & Ground-Truth & Result \\
        \midrule
        Case 1 & 323.9 MB & 3.70sec & true & true \\
        \midrule
        Case 2 & 328.5 MB & 3.59sec & true & true\\  
        \midrule
        Case 3 & 326.4 MB & 3.18sec & false & false \\
\bottomrule
		\label{tab:results}
\end{tabularx}
\end{table}

The model checking results\footnote{See folder `reports' in our online appendix\cite{appendix:online}} return either \emph{Error found} or \emph{No error found}. In our case, \emph{Error found} translates to \emph{true} in \cref{tab:results}, and the other way around for \emph{Error found}. Such translation results from how checking liveness in DIVINE 4 and how we have implemented the Stability property.

\section{Related Work}\label{sec:rw}
\noindent

\one{Among the related literature on applying control theory to self-adaptive systems, there is also some body of work coming from the "aggregate computing" topic that you might want to consider. The topic is surveyed in [1], and one control theory-based paper in this area is for example [2].}
\one{[1] Viroli, M., Beal, J., Damiani, F., Audrito, G., Casadei, R., \& Pianini, D. (2019). From distributed coordination to field calculus and aggregate computing. Journal of Logical and Algebraic Methods in Programming, 109, 100486.[2] Mo, Y., Audrito, G., Dasgupta, S., \& Beal, J. (2020). A Resilient Leader Election Algorithm Using Aggregate Computing Blocks. IFAC-PapersOnLine, 53(2), 3336-3341.}

We start with related work that discusses properties of self-adaptive systems from a SE perspective. Then we discuss related work that emphasizes the importance of relating traditional SE properties to CT properties. 

Back in 2012, the authors of~\cite{weyns:2012:survey} performed a systematic literature review on the use of formal methods in self-adaptive systems. The results show that safety, liveness, and reachability are the main properties considered and these properties are primarily used to verify the efficiency/performance, reliability, and functionality of self-adaptive systems. While instrumental for SE properties considered in self-adaptive systems, the authors do not look into a mapping with CT properties. 

The community papers  \cite{lemos:2017:assurance:challenges} and  \cite{weyns:2019:perpetual} that emerged from a Dagstuhl seminar state that assuring that a self-adaptive system complies with its requirements requires an enduring process that spans the whole lifetime of the system. The authors refer to this process as ``perpetual assurances'' and emphasize that control theory offers a basis to design solutions that provide such assurances. 

The study in~\cite{villegas:2011:framework} is a  pioneering work mapping between quality attributes of self-adaptive systems and properties of control theory. For instance, performance with latency and throughput is mapped to settling time. The proposed mapping is done based on terminology derived from a literature survey. In~\cite{shevtsov:2018:litrev}, the authors derived a mapping between software qualities and control properties based on the results of a systematic literature review on control-theoretical software adaptation. For instance, guaranteeing settling time may be associated with most software qualities since the property refers to guarantees on the time it takes to bring measured quality property close to its goal. In both papers, the presented mapping stayed at a high level in comparison to our work. 

The most relevant paper for the work presented in this paper is~\cite{camara:2020:bridging}. In that paper, the authors mapped key properties that characterize self-adaptive systems to control properties, leveraging the formalization of both in temporal logic. While that work relies on a general formal notation to identify the mapping between SE and CT properties, our work leverages on the structures of a set of established patterns. 

The paper\cite{viroli:2019:aggregate:survey} highlights recent efforts on self-stabilization in aggregate computing, for instance \cite{dasgupta:2016:aggregate,viroli:2018:aggregate,mo:2020:aggregate}. These efforts focus on providing guarantees for control-based properties to algorithms for self-organization of distributed systems, in contrast to mapping how properties are considered in both SE and CT. Furthermore, such effort reinforces the promising path of using of CT tooling for safety assurance provision for aggregate algorithms.
\section{Conclusion and Future Work}
\label{sec:conclusion}
\noindent

\two{could they discuss more about how they can set time constraints (from CT and SE viewpoints) so that the guarantees include the ability to react quickly enough under different anomalous conditions?}\ric{this is already in our future work section.}

To improve the engineering of self-adaptive systems, we proposed a technique to unify the properties used in CT design and those used in SE verification. Our technique relies on using existing specification patterns as a common notation. To that end, we follow a bottom-up approach inspired by Scuderia Ferrari to provide evidence for safety requirements from both CT and SE viewpoints. The properties are mapped using property specification patterns as a common language. Such properties are formalized in LTL and fed into DIVINE 4 to consolidate the proposed mapping through model checking. 
Our initial results are promising based on the mapping.

Future work should identify and map further properties and extend the exploration space of our validation. Our systematic approach for validation of our mapping provides a pathway for building evidence that the mapping is sound. Specifically, we believe the CT properties settling time, overshoot, steady-state error should be mapped to the properties widely used for verification for self-adaptation, e.g., reachability, security, privacy, availability. 
In this work, we considered a CT property that does not require explicit time and for this reason it was enough to map it to untimed property specification patterns, i.e. those initially proposed in~\cite{dwyer:1999:patterns}. It is worth mentioning that there exist also properties specification patterns with explicit time as well as with probability~\cite{autili:2015:aligning}, which might be exploited for other CT properties, when needed. It would be also interesting to extend, if needed, the catalog of specification patterns with specific patterns that are tailored to CT properties. We believe that the mapping could be an important contribution to both the SE and CT communities, by giving, on one side, a concrete instrument for engineering trustworthy and safe autonomous system, and, on the other side, it might facilitate the cross-fertilisation among the two communities. Another direction for future research concerns the investigation of the topics of this paper with companies producing autonomous systems involving both CT controllers and software produced by developers.

\section*{Acknowledgment}
\raz{please check if anything more to be added or taken out}
This work is supported by the Wallenberg AI, Autonomous Systems and Software Program (WASP) funded by the Knut and Alice Wallenberg Foundation. This work is also partly supported by the Swedish Research Council (VR) via the ``PSI'' project. 
The authors also acknowledge financial support from Centre of EXcellence on Connected, Geo-Localized and Cybersecure Vehicle (EX-Emerge), funded by Italian Government under CIPE resolution n. 70/2017 (Aug. 7, 2017).


\balance

\bibliographystyle{IEEEtran}
\bibliography{main}

\end{document}